\documentclass[aps,preprint]{revtex4}%
\usepackage{amsfonts}
\usepackage{amsmath}
\usepackage{amssymb}
\usepackage{graphicx}%
\usepackage{xcolor}
\providecommand{\U}[1]{\protect\rule{.1in}{.1in}}
\newcommand{\nobracket}{}

\begin{document}
\title{Non-Hermitian Noncommutative Quantum Mechanics}
\author{J. F. dos Santos$^{\ast}$, F. S. Luiz$^{\dagger}$, O. S. Duarte$^{\dagger}$
and M. H. Y. Moussa$^{\dagger}$}
\affiliation{$^{\ast}$Universidade Federal do ABC, Santo Andr\'{e}, 09210-580, S\~{a}o
Paulo, Brazil}
\affiliation{$^{\dagger}$Instituto de F\'{\i}sica de S\~{a}o Carlos, Universidade de
S\~{a}o Paulo, P.O. Box 369, S\~{a}o Carlos, 13560-970, S\~{a}o Paulo, Brazil}

\begin{abstract}
In this work we present a general formalism to treat non-Hermitian and
noncommutative Hamiltonians. This is done employing the phase-space formalism of quantum mechanics, which allows to write a set of robust maps connecting
the Hamitonians and the associated Wigner functions to the different Hilbert
space structures, namely, those describing the non-Hermitian and
noncommutative, Hermitian and noncommutative, and Hermitian and commutative
systems. A general recipe is provided to obtain the expected values of the
more general Hamiltonian. Finally, we apply our method to the harmonic
oscillator under linear amplification and discuss the implications of both
non-Hermitian and noncommutative effects.

\end{abstract}
\email{jonas@df.ufscar.br}

\pacs{32.80.-t, 42.50.Ct, 42.50.Dv}
\maketitle

\section{Introduction}

In the last two decades two distinct formalisms have been separately developed
for the treatments of non-Hermitian \cite{Mostafazadeh} or noncommutative
\cite{SW} Hamiltonians. These formalisms extend significantly the scope of
conventional quantum mechanics, whose observables comprise only Hermitian
operators and whose components of specific operators as position and momentum,
always commute with each other. The replacement of the (predominantly
mathematical) requirement of Hermiticity by that (of greater physical bias) of
invariance by spatial reflection and time reversal, enabled the $\mathcal{PT}%
$-symmetric quantum mechanics to contemplate a wide range of physical models
unrelated to Hermitian quantum mechanics \cite{Grounds}. The same occurs when
we impose noncommutation relations between the components of an operator that
were previouly commuting, and new phenomena emerge both from non-Hermitian
\cite{Bender} and noncommutative quantum mechanics \cite{Gouba01, Catarina01}.

There is a strong research activity in the field of non-Hermitian quantum
mechanics, both as regards its foundation \cite{TDNHQM,SCarlos} and its
application to a wide variety of phenomena, such as for metrology
\cite{Metrology}, chaos in optomechanics \cite{Chaos, Chaos02}, and for
stimulating the fluctuation superconductivity \cite{Super}. The finding of
$\mathcal{PT}$-symmetry breaking leading to the coalescence of the energy
levels has also been reported in distinct systems as microwave billiard
\cite{Bittner}, tunneling heterostructures \cite{TH}, lattice model \cite{LM},
a ferromagnetic superfluid \cite{FS}, etc., considerably broadening the range
of interest in the physics of non-Hermitian Hamiltonians.

Noncommutative Hamiltonians, i. e., Hamiltonians described in terms of variables obeying general commutation relations \cite{new03}, have also been useful to study new fundamental
properties emerging when considering a deformed Heisenberg-Weyl algebra
\cite{Snyder}. As examples of this effort, the noncommutative quantum
mechanics has been studied in the harmonic oscillator Hamiltonian
\cite{Rosembaum, Smailagic01, Smailagic02}, in the context of the
thermodynamical limit and quantum information theory \cite{Bernardini01}, in
quantum heat engines \cite{Santos}, and in the well known gravitational
quantum well for ultra-cold neutrons, where corrections in the eigenenergies
have been obtained \cite{Orfeu01, Banerjee}.

In what follows we briefly revisit the foundations of non-Hermitian and
noncommutative quantum mechanics, discussing their key ingredients: the Dyson
map and the associated metric operator for non-Hermitian Hamiltonians, and the
Seiberg-Witten transformation for the noncommutative ones. The Dyson map enable
us to construct the Hermitian counterpart of a non-Hermitian Hamiltonian, both
being isospectral partners, and to define the metric operator which ensures
the unitarity of the time evolution of the system. The Seiberg-Witten
tranformation \cite{SW}, by its turn, allow us to describe a noncommutative
operator in terms of its commutative counterparts, converting all
noncommutative structure to the standard Hilbert space. Although the
Seiberg-Witten map is not unique, once it depends on arbitrary parameters, it
was demonstrated that the eigenvalues of a noncommutative system, such as the
eigenenergies, do not depend on the choice of the map \cite{Catarina01}.

This work is organized as follows. The section \ref{sec2} is devoted to review the
main structures of non-Hermitian and noncommutative quantum mechanics
separately, in order to establish the basic formalism to be used. In section
\ref{sec3} the general non-Hermitian and noncommutative formalism is presented
in detail and the main aspects are discussed. We apply our method to an
illustrative example in section \ref{sec4}. The conclusion and final remarks
are presented in section \ref{con}.

\section{Review of Non-Hermitian and Noncommutative Quantum Mechanics}

\label{sec2}

In this section we consider the main features of the two extensions of quantum
mechanics ---the non-Hermitian and the noncommutative quantum mechanics ---
separately. Concerning the first one, we focus on the Dyson map which enables
the derivation of a Hermitian counterpart of the pseudo-Hermitian
Hamiltonian, whereas the second one is based on the Seiberg-Witten map. Both
maps are described in details in the following.

\subsection*{Non-Hermitian Quantum Mechanics}

Let us begin by revisiting the non-Hermitian quantum mechanics when time
independent Hamiltonians and Dyson maps are considered. As antecipated above,
the key ingredients here are the positive-definite Dyson map $\eta$ and its
associated metric operator $\Theta=\eta^{\dag}\eta$, which ensure,
respectively, the essential features of quantum mechanics: real energy
spectrum and unitary (probability-preserving) time evolution. Let us consider
a non-Hermitian Hamiltonian, $\mathcal{H}_{NH}(\mathbf{q},\mathbf{p})$, which
is $\mathcal{PT}$-symmetric and thus remains invariant under the
transformation%
\begin{equation}
\mathbf{q}\rightarrow-\mathbf{q},\text{ }\mathbf{p}\rightarrow\mathbf{p}%
,\text{ }i\rightarrow-i, \label{q1}%
\end{equation}
Under the quasi-Hermiticity relation%
\begin{equation}
\Theta(\mathbf{q},\mathbf{p})\mathcal{H}_{NH}(\mathbf{q},\mathbf{p}%
)=\mathcal{H}_{NH}^{\dagger}(\mathbf{q},\mathbf{p})\Theta(\mathbf{q}%
,\mathbf{p}), \label{eq2}%
\end{equation}
the non-Hermitian $\mathcal{H}_{NH}(\mathbf{q},\mathbf{p})$ is mapped to its
Hermitian counterpart
\begin{equation}
H_{H}(\mathbf{q},\mathbf{p})=\eta(\mathbf{q},\mathbf{p})\mathcal{H}%
_{NH}(\mathbf{q},\mathbf{p})\eta^{-1}(\mathbf{q},\mathbf{p}), \label{eq3}%
\end{equation}
where the Dyson map as well as the metric operator is defined by the same set
of operators as the Hamiltonians.

From the solutions of the Schr\"{o}dinger equations for $\mathcal{H}_{NH}$ and
$H_{H}$, given by $\left\{  \left\vert \psi_{n}\right\rangle \right\}  $ and
$\left\{  \left\vert \varphi_{n}\right\rangle \right\}  $, respectively, with
$\left\vert \varphi_{n}\right\rangle =\eta\left\vert \psi_{n}\right\rangle $,
we immediately verify that both Hamiltonians are isospectral partners, sharing
the same eigenvalues $\left\{  \varepsilon_{n}\right\}  $. Moreover, its is
straightforward to see that, besides having real energy spectrum, the
non-Hermitian $\mathcal{H}_{NH}$ generates unitary time evolution under the
redefined metric
\begin{equation}
\left\langle \Psi\left\vert \tilde{\Psi}\right.  \right\rangle _{\Theta}%
\equiv\left\langle \Psi\left\vert \Theta\tilde{\Psi}\right.  \right\rangle
=\left\langle \Phi\left\vert \tilde{\Phi}\right.  \right\rangle , \label{eq4}%
\end{equation}
the capital $\Psi$ and $\tilde{\Psi}$, (as well as $\Phi$ and $\tilde{\Phi}$)
being generic superpositions of the basis states $\left\{  \left\vert \psi
_{n}\right\rangle \right\}  $ ($\left\{  \left\vert \varphi_{n}\right\rangle
\right\}  $). The matrix elements of the observables,
\begin{equation}
\mathcal{O}(\mathbf{q},\mathbf{p})=\eta^{-1}(\mathbf{q},\mathbf{p}%
)O(\mathbf{q},\mathbf{p})\eta(\mathbf{q},\mathbf{p}) \label{eq5}%
\end{equation}
associated with the non-Hermitian $H$ are accordingly computed in the new
metric as
\begin{equation}
\left\langle \psi_{m}\left\vert \mathcal{O}\right\vert \psi_{n}\right\rangle
_{\Theta}=\left\langle \psi_{m}\left\vert \Theta\mathcal{O}\right\vert
\psi_{n}\right\rangle =\left\langle \varphi_{m}\left\vert O\right\vert
\varphi_{n}\right\rangle , \label{eq6}%
\end{equation}
the calligraphyc (italic) capitals referring to the non-Hermitian (Hermitian)
system, as it has become clear by now.

Starting from the density operator $\rho_{\varphi}=\sum\nolimits_{n}%
P_{n}\left\vert \varphi_{n}\right\rangle \left\langle \varphi_{n}\right\vert $
associated with the Hermitian $H_{H}$,  and using the biorthonormal basis
$\left\{  \left\vert \psi_{n}\right\rangle ,\left\vert \upsilon_{n}%
\right\rangle \right\}$ \cite{Brody2014}, where $\left\vert \upsilon_{n}\right\rangle
=\eta^{\dag}\left\vert \varphi_{n}\right\rangle $ is an eigenvector of
$\mathcal{H}_{NH}^{\dag}$ and $| \psi_n \rangle  = \eta^{-1}|\varphi_n\rangle$ is an eigenvector of $\mathcal{H}_{NH}$, it follows immediately that the density operator
associated with the non-Hermitian $\mathcal{H}_{NH}$ is given by $\rho_{\psi
}=\eta^{-1}\rho_{\varphi}\eta=\sum\nolimits_{n}P_{n}\left\vert \psi
_{n}\right\rangle \left\langle \psi_{n}\right\vert $, and the
quasi-Hermiticity relation $\Theta\rho_{\psi}=$ $\rho_{\psi}^{\dag}\Theta$ is
satisfied. It is straightforward to verify that $\operatorname*{Tr}\left(
\Theta\rho_{\psi}\right)  =\sum\nolimits_{n}\left\langle \psi_{n}\right\vert
\Theta\rho_{\psi}\left\vert \psi_{n}\right\rangle =\sum\nolimits_{n}P_{n}=1$
and the expected values,
\begin{equation}
\left\langle \mathcal{O}\right\rangle _{\Theta}=\operatorname{Tr}\left(
\rho_{\psi}\mathcal{O}\right)  _{\Theta}=\sum\nolimits_{n}P_{n}\left\langle
\psi_{n}\right\vert \Theta\mathcal{O}\left\vert \psi_{n}\right\rangle
=\operatorname{Tr}[\rho_{\varphi}O]=\left\langle O\right\rangle .\label{eq7}%
\end{equation}

The expected value from an observable obtained from the density operator
formalism will be very useful when dealing with the Wigner function and its
map from the non-Hermitian and noncommutative Hamiltonian to the Hermitian and
commutative one.

\subsection*{Noncommutative Quantum Mechanics}

Regarding noncommutative quantum mechanics, deformed commutation relations are
defined under the assumption that a different space-time structure emerges in
the Planck scale, $l_{P}=10^{-33}cm$ \cite{SW, Douglas,Wei,Balachandran01}:
\begin{equation}
\lbrack q_{k},q_{\ell}]=i\theta_{k\ell},\text{ \ }[q_{k},p_{\ell}%
]=i\hbar\delta_{k\ell},\text{ \ }[p_{k},p_{\ell}]=i\zeta_{k\ell}, \label{2}%
\end{equation}
with $k$ and $\ell$ labeling the components of the position and momentum
operators, whereas the quantities $\theta_{k\ell}=\theta\epsilon_{k\ell}$ and
$\zeta_{k\ell}=\zeta\epsilon_{k\ell}$ introduce two new fundamental constants
in the theory, $\theta$ and $\zeta$, with units $[\theta]=L^{2}$ and
$[\zeta]=M^{2}L^{2}/T^{2}$. Here, $\epsilon_{kk}=0$ and $\epsilon_{k\ell
}=-\epsilon_{\ell k}$, in contrast with the standard commutation relations,
\begin{equation}
\lbrack Q_{k},Q_{\ell}]=0,[Q_{k},P_{\ell}]=i\hbar\delta_{k\ell},[P_{k}%
,P_{\ell}]=0. \label{3}%
\end{equation}

The connection between the noncommutative algebra and the standard one is
performed through the Seiberg-Witten map \cite{SW}, which is a linear
transformation written in general as $q_{k}=A_{k\ell}Q_{\ell}+B_{k\ell}%
P_{\ell},$ $p_{k}=C_{k\ell}Q_{\ell}+D_{k\ell}P_{\ell},$ where $A_{k\ell}$,
$B_{k\ell}$, $C_{k\ell}$ and $D_{k\ell}$ are elements of the matrices
$\mathbf{A}$, $\mathbf{B}$, $\mathbf{C}$, and $\mathbf{D}$, assumed reals,
constants and invertibles. Using the relations (\ref{2}) and (\ref{3}) we
derive Seiberg-Witten map,
\begin{equation}
q_{k}=\nu Q_{k}-\frac{\theta_{k\ell}}{2\nu\hbar}P_{\ell},\text{ \ \ }p_{k}=\mu
P_{k}+\frac{\zeta_{k\ell}}{2\mu\hbar}Q_{\ell}, \label{4}%
\end{equation}
with $\mu$ and $\nu$ being arbitrary parameters constrained by
\begin{equation}
\frac{\theta\zeta}{4\hbar^{2}}=\nu\mu(1-\nu\mu). \label{5}%
\end{equation}
The inverse Seiberg-Witten map reads,
\begin{subequations}
\label{6}%
\begin{align}
Q_{k}  &  =\mu\left(  1-\frac{\theta\zeta}{\hbar^{2}}\right)  ^{-1/2}\left(
q_{k}+\frac{\theta_{k\ell}}{2\nu\mu\hbar}p_{\ell}\right)  ,\label{6a}\\
P_{k}  &  =\nu\left(  1-\frac{\theta\zeta}{\hbar^{2}}\right)  ^{-1/2}\left(
p_{k}-\frac{\zeta_{k\ell}}{2\nu\mu\hbar}q_{\ell}\right)  . \label{6b}%
\end{align}

From equations (\ref{6a}) and (\ref{6b}) one notices that the commutative variables are recovered when $(\theta, \zeta) \rightarrow(0,0)$. When we look at the equations (\ref{6a}) and (\ref{6b}) more carefully, we see that there is apparently a critical point when $\zeta = \hbar^{2}/\theta$. In order to avoid this critical point it is necessary to impose  $\theta\zeta\neq\hbar^{2}$, thereby ensuring that the Seiberg-Witten map is invertible \cite{Catarina01}. Besides, the way how the commutative variables are written as function of the noncommutative ones is responsible for the NC effects quantified by the noncommutative constants once the Hamiltonian of the system is defined.

\section{Non-Hermitian NonCommutative Quantum Mechanics}
\label{sec3}

In order to put together the non-Hermitian and the noncommutative formalisms,
we first observe that the Seiberg-Witten is not a $\mathcal{PT}$-symmetric
map. Therefore, when starting from a non-Hermitian noncommutative
$\mathcal{PT}$-symmetric Hamiltonian $\mathcal{H}_{NH,NC}^{\mathcal{PT}%
}(\mathbf{q},\mathbf{p})$, we can not first employ the Seiberg-Witten map to
transform this Hamiltonian to a commutative non-Hermitian one, since in this
way we lose its fundamental $\mathcal{PT}$-symmetric character, and we are
left with a non-physical (non-Hermitian non-$\mathcal{PT}$-symmetric)
Hamiltonian. Therefore, the only way to bring both formalisms together is to
first transform the Hamiltonian $\mathcal{H}_{NH,NC}^{\mathcal{PT}}%
(\mathbf{q},\mathbf{p})$, through the Dyson map, into the Hermitian
noncommutative one $H_{H,NC}(\mathbf{q},\mathbf{p})$, and then to use the
Seiberg-Witten map to reach the Hermitian commutative form $\mathrm{H}%
_{H,C}(\mathbf{Q},\mathbf{P})$, i.e.,
\end{subequations}
\begin{equation}
\mathcal{H}_{NH,NC}^{\mathcal{PT}}(\mathbf{q},\mathbf{p})\overset
{\eta(\mathbf{q},\mathbf{p})}{\rightarrow}H_{H,NC}(\mathbf{q},\mathbf{p}%
)\overset{Seiberg-Witten}{\rightarrow}\mathrm{H}_{H,C}(\mathbf{Q},\mathbf{P}),
\label{7}%
\end{equation}
where, referring to the non-Hermitian noncommutative, Hermitian
noncommutative and Hermitan commutative systems, respectively, we use
calligraphic, italic and roman capitals letters.

As a next step, we will introduce another well known way to describe the state
of a quantum system, the so called Wigner function \cite{Wigner}, which has
been extensively used to describe noncommutative quantum systems. This is done
because, unlike the wave function, the Wigner function contains information
about both position and momentum and then it is possible to describe
noncommutative states. To obtain the Wigner function, we first introduce the
Weyl transform which converts an operator in a c-number, defined as
\cite{Catarina01},
\begin{equation}
\mathrm{A}^{W}(\mathbf{Q},\mathbf{P})=\int d\mathbf{y}\,e^{-i\mathbf{P}%
.\mathbf{y}/\hbar}\langle\mathbf{Q}-\mathbf{y}/2|\mathrm{A}(\mathbf{Q}%
,\mathbf{P})|\mathbf{Q}+\mathbf{y}/2\rangle\text{,} \label{eq8}%
\end{equation}
where $\mathrm{A}(\mathbf{Q},\mathbf{P})$ is an arbitrary operator and the
superscript ``W" stands for the Weyl transform. The Wigner function is the
Weyl transform of the density operator and, given the Hamiltonian
$\mathrm{H}_{H,C}(\mathbf{Q},\mathbf{P})$ and the associated density operator
$\rho$, it follows that%

\begin{equation}
\mathrm{W}(\mathbf{Q},\mathbf{P})=\int d\mathbf{y}e^{-i\mathbf{P}%
\cdot\mathbf{y}/\hbar}\langle\mathbf{Q}+\mathbf{y}/2|\rho|\mathbf{Q}%
-\mathbf{y}/2\rangle, \label{eq9}%
\end{equation}
with the probability density in position (momentum) representation being
obtained by integrating the Wigner function over the momentum (position)
variable. The Wigner-Weyl formalism of quantum mechanics also encompasses the
so called Moyal product \cite{Catarina01,Zachos}, which allow us to obtain all
features of the quantum mechanics formalism based in operator structure and is
defined as,
\begin{equation}
\star=\star_{\hbar}\star_{\theta}\star_{\zeta} \label{eq12}%
\end{equation}
where each term, given by
\begin{subequations}
\label{eq13}%
\begin{align}
\star_{\hbar}  &  =\sum_{k}\exp\left[  \frac{i\hbar}{2}\left(  \frac
{\overleftarrow{\partial}}{\partial q_{k}}\frac{\overrightarrow{\partial}%
}{\partial p_{k}}-\frac{\overleftarrow{\partial}}{\partial p_{k}}%
\frac{\overrightarrow{\partial}}{\partial q_{k}}\right)  \right]
,\label{eq13a}\\
\star_{\theta}  &  =\sum_{k,\ell,k\neq\ell}\exp\left[  \frac{i\theta}{2}\left(
\frac{\overleftarrow{\partial}}{\partial q_{k}}\frac{\overrightarrow{\partial
}}{\partial q_{\ell}}-\frac{\overleftarrow{\partial}}{\partial q_{\ell}}%
\frac{\overrightarrow{\partial}}{\partial q_{k}}\right)  \right]
,\label{eq13b}\\
\star_{\zeta}  &  =\sum_{k,\ell,k\neq\ell}\exp\left[  \frac{i\zeta}{2}\left(
\frac{\overleftarrow{\partial}}{\partial p_{k}}\frac{\overrightarrow{\partial
}}{\partial p_{\ell}}-\frac{\overleftarrow{\partial}}{\partial p_{\ell}}%
\frac{\overrightarrow{\partial}}{\partial p_{k}}\right)  \right]  ,
\label{eq13c}%
\end{align}
refers to an associated commutation relation in eq.(\ref{2}), with the
arrows, pointing to the right and left, indicating the directions of operation
of the derivatives.  The star-product $\star$ clearly reduces to $\star_\hbar$ when we are into the Hermitian commutative quantum mechanics. This product is very suitable when the phase-space formalism of quantum mechanics is employed, because there is no more operators and every physical quantity is represented by c-functions which, using the standard inner product, commute each other. By using the star-product we replace the commutation of operators by partial derivative of position and momentum, and all the relevant results can be recovered. Naturally, the general star-product $\star$ is the extension for encompassing the noncommutative algebra given by eq.(\ref{2}). For a more detailed discussion we addressed the ref. \cite{Catarina01}.

An alternative form to compute the Wigner function $\mathrm{W}(\mathbf{Q}%
,\mathbf{P})$ given in eq.(\ref{eq9}) follows by applying the Weyl transform
on the eigenvalue equation for $\mathrm{H}_{H,C}(\mathbf{Q},\mathbf{P})$ in
the density matrix formulation, i.e., $\mathrm{H}_{H,C}(\mathbf{Q}%
,\mathbf{P})\left\vert E\right\rangle \left\langle E\right\vert =E$
$\left\vert E\right\rangle \left\langle E\right\vert $, what leads to the
$\star$-value equation \cite{Rosembaum},
\end{subequations}
\begin{equation}
\mathrm{H}_{H,C}^{W}(\mathbf{Q},\mathbf{P})\star\mathrm{W}(\mathbf{Q}
,\mathbf{P})=E\,\mathrm{W}(\mathbf{Q},\mathbf{P}), \label{eq14}%
\end{equation}
where $\mathrm{H}_{H,C}^W$ represents the Weyl transform of the associated
Hamiltonian, $\mathrm{W}(\mathbf{Q},\mathbf{P})$ describes the state of the
system and $E$ are the eigenenergies.

In order to unify the notation, we refer to $\rho$, $\rho_{\varphi}$, and
$\rho_{\psi}$ as being the density operators related to the Hamiltonians
$\mathrm{H}_{H,C}$, $H_{H,NC}$, and $\mathcal{H}_{NH,NC}^{\mathcal{PT}}$,
respectively, as well as the definition introduced in eq.(\ref{7}) applied to
the Wigner functions [$\mathrm{W}(\mathbf{Q},\mathbf{P})$, $W(\mathbf{q}%
,\mathbf{p})$ and $\mathcal{W}(\mathbf{q},\mathbf{p})$] and the observables
[$\mathrm{O}(\mathbf{Q},\mathbf{P})$, $O(\mathbf{q},\mathbf{p})$ and
$\mathcal{O}(\mathbf{q},\mathbf{p})$]. Using the inverse Seiberg-Witten map,
we straightfowardly obtain the Wigner function $W(\mathbf{q},\mathbf{p})$ from
$\mathrm{W}(\mathbf{Q},\mathbf{P})$ \textbf{\cite{Catarina01}}, and by
Weyl-transforming the expression $\rho_{\varphi}=\eta\rho_{\psi}\eta^{-1}$, we
automatically derive a relation between $W(\mathbf{q},\mathbf{p})$ and
$\mathcal{W}(\mathbf{q},\mathbf{p})$; from $\rho_{\varphi}^{W}=\left(
\eta\rho_{\psi}\eta^{-1}\right)  ^{W}=\eta^{W}\star\rho_{\psi}^{W}\star\left(
\eta^{-1}\right)  ^{W}$ we obtain
\begin{equation}
W(\mathbf{q},\mathbf{p})=\eta^{W}(\mathbf{q},\mathbf{p})\star\mathcal{W}%
(\mathbf{q},\mathbf{p})\star\left(  \eta^{-1}\right)  ^{W}(\mathbf{q}%
,\mathbf{p})\text{,}\label{eq10}%
\end{equation}
or equivalently
\begin{equation}
\mathcal{W}(\mathbf{q},\mathbf{p})=\left(  \eta^{-1}\right)  ^{W}%
(\mathbf{q},\mathbf{p})\star W(\mathbf{q},\mathbf{p})\star\left(  \eta\right)
^{W}(\mathbf{q},\mathbf{p}).\label{eq11}%
\end{equation}

Note that the star product given by eq.(\ref{eq12}) was essential to obtain a
phase-space representation of the map connecting the non-Hermitian to the
Hermitan counterparts. Just for stress the results above, $\mathcal{W}%
(\mathbf{q},\mathbf{p})$ is the Wigner function associated with the
non-Hermitian noncommutative Hamiltonian whereas $\mathrm{W}(\mathbf{Q}%
,\mathbf{P})$ is the counterpart associated to the Hermitian and commutative
case. In the following we elucidate how the description of the map links the
observables from NHNC Hamiltonians to the HC ones.

\subsection*{Expected Values}

Starting again from the Hermitian commutative Hamiltonian, the expectation
value of an observable $\mathrm{O}(\mathbf{Q},\mathbf{P})$ associated with
$\mathrm{H}_{H,C}^{W}(\mathbf{Q},\mathbf{P})$, in the phase-space formalism of
quantum mechanics, is given by \cite{Case}:
\begin{equation}
\langle\mathrm{O}(\mathbf{Q},\mathbf{P})\rangle=\int\int d\mathbf{Q}%
d\mathbf{P}\,\mathrm{W}(\mathbf{Q},\mathbf{P})\mathrm{O}^{W}(\mathbf{Q}%
,\mathbf{P}). \label{eq15}%
\end{equation}
Using the inverse Seiberg-Witten map, we verify that the expectation value
$\langle\mathrm{O}(\mathbf{Q},\mathbf{P})\rangle$ is related to that
associated with the Hamiltonian $H_{H,NC}\left(  \mathbf{q},\mathbf{p}\right)
$ in the form%

\begin{align}
\langle\mathrm{O}(\mathbf{Q},\mathbf{P})\rangle &  =\int\int d\mathbf{q}%
d\mathbf{p}\frac{\partial(\mathbf{Q},\mathbf{P})}{\partial(\mathbf{q}%
,\mathbf{p})}\mathrm{W}(\mathbf{Q}(q, p),\mathbf{P}(q, p))\mathrm{O}%
^{W}(\mathbf{Q}(q, p),\mathbf{P}(q, p))\nonumber\\
&  =\int\int d\mathbf{q}d\mathbf{p}W(\mathbf{q},\mathbf{p})O^{W}%
(\mathbf{q},\mathbf{p})=\langle O(\mathbf{q},\mathbf{p})\rangle,\label{eq16}%
\end{align}
$\partial(\mathbf{Q},\mathbf{P})/\partial(\mathbf{q},\mathbf{p})$ being the
Jacobian of the transformation from $(\mathbf{Q},\mathbf{P})$ to
$(\mathbf{q},\mathbf{p})$, and the Wigner function $W(\mathbf{q},\mathbf{p})$
associated with the Hermitian noncommutative Hamiltonian $H_{H,NC}%
(\mathbf{q},\mathbf{p})$ is defined as \cite{Catarina01}
\begin{equation}
W(\mathbf{q},\mathbf{p}) = \frac{\partial(\mathbf{Q},\mathbf{P})}%
{\partial(\mathbf{q},\mathbf{p})}\mathrm{W}(\mathbf{Q}(q, p),\mathbf{P}(q,
p)).
\end{equation}

Now, to compute the expectation value of the observable $\mathcal{O}%
(\mathbf{Q},\mathbf{P})$ we redefine the metric to obtain
\begin{equation}
\langle\mathcal{O}(\mathbf{q},\mathbf{p})\rangle_{\Theta}=\int\int
d\mathbf{q}d\mathbf{p}\mathcal{W}(\mathbf{q},\mathbf{p})\Theta^{W}%
(\mathbf{q},\mathbf{p})\mathcal{O}^{W}(\mathbf{q},\mathbf{p}), \label{eq17}%
\end{equation}
which, together with eq.(\ref{eq11}), the Weyl transform of the metric
$\Theta^{W}=\left(  \eta^{\dagger}\right)  ^{W}\star\eta^{W}$, and the
relation between the Weyl transforms of the observables $\mathcal{O}$ and $O$
(derived using the property $\left(  ABC\right)  ^{W}=A^{W}\star B^{W}\star
C^{W}$:
\begin{equation}
\mathcal{O}^{W}(\mathbf{q},\mathbf{p})=\left(  \eta^{-1}\right)
^{W}(\mathbf{q},\mathbf{p})\star O^{W}(\mathbf{q},\mathbf{p})\star\eta
^{W}(\mathbf{q},\mathbf{p}). \label{eq18}%
\end{equation}
enable us to verify, as required, that the expectation value $\langle
\mathcal{O}(\mathbf{q},\mathbf{p})\rangle_{\Theta}$ equals $\left\langle
O(\mathbf{q},\mathbf{p})\right\rangle $ and $\langle\mathrm{O}(\mathbf{Q}%
,\mathbf{P})\rangle$:
\begin{equation}
\langle\mathcal{O}(\mathbf{q},\mathbf{p})\rangle_{\Theta}=\int\int
d\mathbf{q}d\mathbf{p}W(\mathbf{q},\mathbf{p})O^{W}(\mathbf{q},\mathbf{p}%
)=\langle O(\mathbf{q},\mathbf{p})\rangle=\langle\mathrm{O}(\mathbf{Q}%
,\mathbf{P})\rangle, \label{eq19}%
\end{equation}
where, inside the integrals, the $\star$-product reduces to the regular
operator product \cite{Zachos}. The expression above is our last important
result and it completes the formalism to describe non-Hermitian noncommutative
Hamiltonians by establishing maps connecting the different Hilbert structures
in a robust way.

\section{Illustrative Example: The linearly amplified harmonic oscillator}

\label{sec4}

After the NHNC QM formalism has been constructed, we would like to present an
example. Let us consider the following non-Hermitian and noncommutative
$\mathcal{PT}$-symmetric Hamiltonian,
\begin{equation}
\mathcal{H}_{NH,NC}^{\mathcal{PT}}(\mathbf{q},\mathbf{p})=\sum_{i=1}^{2}
\frac{p_{i}^{2}}{2m}+\frac{1}{2}m\omega_{i}^{2}q_{i}^{2}+\gamma_{i}%
p_{i}+i\delta_{i}q_{i}, \label{11}%
\end{equation}
where $m$, $\omega_{i}$, $\gamma_{i}$, and $\delta_{i}$ are all real constant coefficients and the operators $p$ and $q$ obey the commutation relation (\ref{2}). The Hamiltonian (\ref{11}) becomes Hermitian only if $\delta_{i}=0$, since all the coefficients are real. As we want to work only with the non-Hermitian Hamiltonian we will impose $\delta_{i}\neq0$. Considering the anzats for the Dyson map
\begin{equation}
\eta(\mathbf{q},\mathbf{p})= \sum_{i=1}^{2}e^{A_{i}\hat{q}_{i}+B_{i}\hat
{p}_{i}}, \label{12}%
\end{equation}
$A_{i}$ and $B_{i}$ being complex coefficients, so we can rewrite these
coefficients in the polar form $A_{i}=|A_{i}|e^{i\theta_{A_{i}}}$ and
$B_{i}=|B_{i}|e^{i\theta_{B_{i}}}$. Using the relation $H_{H,NC}(\mathbf{q},\mathbf{p}) =\eta(\mathbf{q},\mathbf{p})\mathcal{H}_{NH,NC}^{\mathcal{PT}}(\mathbf{q},\mathbf{p})\eta^{-1}(\mathbf{q},\mathbf{p})$, we
derive the Hermitian counterpart of Hamiltonian (\ref{11}):
\begin{equation}
H_{H,NC}( q, p) = \sum_{i = 1}^{2} \left\{  \frac{p_{i}^{2}}{2m} +\frac{1}%
{2}m\omega_{i}^{2}q_{i}^{2} + p_{i} V_{i} + q_{i} T_{i} + \frac{1}%
{2m\omega_{i}^{2}} \left[  m^{2}\omega_{i}^{2} ( V_{i}^{2} - \gamma_{i}^{2}) +
\left(  \delta_{i}^{2} + T_{i}^{2}\right)  \right]  \right\}  ,
\label{fullHHNC}%
\end{equation}
where
\begin{subequations}
\begin{align}
V_{i}  &  = - \frac{\varsigma_{ji} \hbar( \delta_{j} + \delta_{j}^{\ast}) (
\tan[ \theta_{B_{j}}] - \tan[\theta_{A_{i}}])}{2m^{2}\omega_{j}^{2} (
\varsigma_{ji} \theta_{ij} + \hbar^{2})} + \gamma_{i},\\
T_{i}  &  =\frac{i( \delta_{i} - \delta_{i}^{\ast})}{2}+ \frac{( \delta_{i} +
\delta_{i}^{\ast})( \tan[ \theta_{A_{j}}] \theta_{ji}\varsigma_{ij} + \tan[
\theta_{B_{i}}] \hbar^{2})}{2( \theta_{ji} \varsigma_{ij} + \hbar^{2})},
\end{align}
with $j=i+1 \,\text{mod2}$. Note that the coefficients $\delta$ is real in eq.
(\ref{11}) in order to have a non-Hermitian Hamiltonian. However, we will
consider the general case to analyze how this can affect our results.

For the Hamiltonian $H_{H,NC}(q,p)$ to be Hermitian, we need to impose that
the real part of the coefficients in Dyson map are
\end{subequations}
\begin{subequations}
\begin{align}
\mathfrak{R}\left[  A_{i}\right]  =|A_{i}|\cos\left[  \theta_{A_{i}}\right]
&  =-\frac{(\delta_{j}+\delta_{j}^{\ast})\varsigma_{ji}}{2m\omega_{j}%
^{2}(\varsigma_{ji}\theta_{ij}+\hbar^{2})},\\
\mathfrak{R}\left[  B_{i}\right]  =|B_{i}|\cos\left[  \theta_{B_{i}}\right]
&  =\frac{(\delta_{i}+\delta_{i}^{\ast})\hbar}{2m\omega_{i}^{2}(\theta
_{ji}\varsigma_{ij}+\hbar^{2})},
\end{align}
and, on the other hand, there is no restriction on the imaginary part of the
coefficients in the Dyson map. It is evident that the Hamiltonian (\ref{11}) is
$\mathcal{PT}$-symmetric if $\delta_{j}$ is real, resulting in $(\delta
_{j}+\delta_{j}^{\ast})=2\delta_{j}$. However, if we assume the Hamiltonian
(\ref{11}) is Hermitian, this implies $\delta$ must be purely imaginary,
$\delta_{j}=i\tilde{\delta_{j}}$, leading the coefficients of the Dyson map to
zero and, consequently, the Dyson map to unity operator and the Hamiltonian
(\ref{fullHHNC}) becomes the Hermitian form of the Hamiltonian (\ref{11}), as
expected. If we assume coefficients of the Dyson map purely real, and
$\delta_{i}\in\mathbb{R}$, the Hamiltonian $H_{H,NC}(q,p)$ reduces to
\end{subequations}
\begin{equation}
H_{H,NC}(q,p)=\sum_{i=1}^{2}\left[  \frac{p_{i}^{2}}{2m}+\frac{1}{2}%
m\omega_{i}^{2}q_{i}^{2}+\gamma_{i}\,p_{i}+\frac{\delta_{i}^{2}}{2m\omega
_{i}^{2}}\right]  .\label{firstcase}%
\end{equation}
Comparing with the Hamiltonian (\ref{fullHHNC}) is easy to see that the imaginary
part of Dyson map are responsible for the position linear term in the
Hamiltonian (\ref{fullHHNC}). As we note from eq.(\ref{firstcase}), this
simple case leads to a shift in the eigenenergies of the system, which depend
exclusively on the parameters of the Hamiltonian, while in the Hamiltonian
(\ref{fullHHNC}) these terms also depend on the Dyson map parameters. Next,
following the formalism developed here, we have to apply the Seiberg-Witten
map, eq.(\ref{4}), in eq.(\ref{fullHHNC}) resulting in
\begin{align}
H_{H}(Q,P) &  =\sum_{i=1}^{2}\left[  \left(  \frac{\mu^{2}}{2m}+\frac
{m\omega_{j}^{2}\theta_{ji}^{2}}{8\nu^{2}\hbar^{2}}\right)  P_{i}^{2}+\left(
\frac{1}{2}m\omega_{i}^{2}\nu^{2}+\frac{\varsigma_{ji}^{2}}{8m\mu^{2}\hbar
^{2}}\right)  Q_{i}^{2}+\left(  \frac{\varsigma_{ij}-m^{2}\omega_{j}^{2}%
\theta_{ji}}{4m\hbar}\right)  \right.  \nonumber\\
&  \left.  \times\{P_{i},Q_{j}\}+\Xi_{i}P_{i}+\Lambda_{i}Q_{i}+\Omega
_{i}\right.  \Big],\label{HHQP}%
\end{align}
where $\Lambda_{i}=\left(  \frac{\varsigma_{ji}V_{j}}{2\mu\hbar}+\nu
T_{i}\right)  $, $\Xi_{i}=\left(  \mu V_{i}-\frac{\theta_{\operatorname{ji}%
}T_{j}}{2\nu\hbar}\right)  $, $\Omega_{i}=\frac{m^{2}\omega_{i}^{2}(V_{i}%
^{2}-\gamma_{i}^{2})+(T_{i}^{2}+\delta_{i}^{2})}{2m\omega_{i}^{2}}$ and
$\{P_{i},Q_{j}\}=P_{i}Q_{j}+Q_{j}P_{i}$. Unlike the Hamiltonian
(\ref{fullHHNC}), imposing that the Dyson map coefficients are real, and $\delta
\in\mathbb{R}$, none of the terms of the Hamiltonian goes to zero. However, as
the Hamiltonian (\ref{fullHHNC}) when $\{\omega_{i},\gamma_{i}\}\in\mathbb{R}%
$, $\delta=-i\tilde{\delta}$, $\theta_{ij}=\zeta_{ij}\rightarrow0$ and
$\mu=\nu\rightarrow1$, the Hamiltonian (\ref{HHQP}) goes to Hamiltonian
(\ref{11}), as expected. In order to verify the join effects on
non-Hermiticity and non-commutative, we obtain analytically the eigenstates of
the Hamiltonian (\ref{HHQP}). These eigenstates are given by a displacement in
the modes of the number states, followed by a squeezed in those modes and by
two rotation,
\begin{equation}
|\psi\rangle=U(\Upsilon)K(\Gamma)S_{1}(r_{1},\phi_{1})S_{2}(r_{2},\phi
_{2})D_{1}(-\chi_{1})D_{2}(-\chi_{2})|n_{1},n_{2}\rangle,\label{eigenstate}%
\end{equation}
where $n_{i}= 0, 1, 2, \ldots$ are the number states, the unitary operators can be written in
terms of position and moment as follows: $U(\Upsilon)=e^{i\frac{\Upsilon}%
{2}\left(  m\sqrt{\omega_{1}\omega_{2}}Q_{1}Q_{2}+\frac{P_{1}P_{2}}%
{m\sqrt{\omega_{1}\omega_{2}}}\right)  }$, $K(\Gamma)=e^{i\frac{\Gamma}%
{2\hbar}\left(  m\sqrt{\omega_{1}\omega_{2}}Q_{1}Q_{2}-\frac{P_{1}P_{2}%
}{m\sqrt{\omega_{1}\omega_{2}}}\right)  }$, $S_{i}(r_{i},\phi_{i}%
)=e^{-i\frac{r_{i}}{2\hbar}\left(  \sin[\phi_{i}](\frac{P_{i}^{2}}{m\omega
_{i}}-m\omega_{i}Q_{i}^{2})+\cos[\phi_{i}]\{Q_{i},P_{i}\}\right)  }$, and
$D_{i}(\chi_{i})=e^{(\chi_{i}-\chi_{i}^{\ast})\sqrt{\frac{m\omega_{i}}{2\hbar
}}Q_{i}-(\chi_{i}+\chi_{i}^{\ast})\sqrt{\frac{1}{2m\omega_{i}\hbar}}P_{i}}$
with $\phi_{i}=k_{i}\pi$, where $k_{i}=0,1,2\ldots$, $i=1,2$ and,
\[
\Upsilon=\tan^{-1}\left[  \frac{2\hbar\lbrack2F_{-}l_{-}+G_{+}l_{+}]}%
{4(f_{1}^{2}-f_{2}^{2})-(g_{1}^{2}-g_{2}^{2})}\right]  ,\quad\Gamma=\tan
^{-1}\left[  \frac{G_{-}\tan\left[  \Upsilon\right]  +2l\hbar}{2F_{-}%
\sqrt{1+\tan^{2}\left[  \Upsilon\right]  }}\right]  ,
\]%
\begin{align}
\tanh[2r_{1}]  & =\frac{(-1)^{k+1}\left[  2\cosh(\Gamma)F_{-}+2\cos
(\Upsilon)(F_{+}-l_{+}\hbar\sinh[\Gamma])+\sin[\Upsilon](2l_{-}\hbar
-\sinh[\Gamma]G_{-})\right]  }{-2\sin[\Upsilon](\sinh[\Gamma]F_{+}%
+\hbar)+\cosh[\Gamma]G_{+}+\cos[\Upsilon](G_{-}+2l_{-}\hbar\sinh[\Gamma])},\\
\tanh[2r_{2}]  & =-\frac{(-1)^{k+1}\left[  2\cosh[\Gamma]F_{-}-2\cos
[\Upsilon](F_{+}+l_{+}\hbar\sinh[\Gamma])-\sin[\Upsilon](\sinh[\Gamma
]G_{-}+2l_{-}\hbar)\right]  }{2\sinh[\Gamma](l_{-}\hbar\cos[\Upsilon
]-F_{+}\sin[\Upsilon])+\cosh[\Gamma](-G_{-}\cos[\Upsilon]+2\hbar\sin
[\Upsilon])+G_{+}},
\end{align}%
\begin{equation}
\chi_{1}=\frac{\sqrt{2}\cosh[2r_{1}]\left(  i\Xi_{1}\kappa_{1_{+}}%
\sqrt{m\omega_{1}\hbar}+\Xi_{2}\lambda_{1_{+}}\sqrt{m\omega_{2}\hbar}%
+\Lambda_{1}\kappa_{1_{-}}\sqrt{\frac{\hbar}{m\omega_{1}}}-i\Lambda_{2}%
\lambda_{1_{-}}\sqrt{\frac{\hbar}{m\omega_{2}}}\right)  }{-2\sin
[\Upsilon](\sinh[\Gamma]F_{+}+l_{+}\hbar)+\cosh[\Gamma]G_{+}+\cos
[\Upsilon](G_{-}+2l_{-}\hbar\sinh[\Gamma])},\label{xi1}%
\end{equation}%
\begin{equation}
\chi_{2}=\frac{\sqrt{2}\cosh[2r_{2}]\left(  \Xi_{1}\lambda_{2_{+}}%
\sqrt{m\omega_{1}\hbar}+i\Xi_{2}\kappa_{2_{+}}\sqrt{m\omega_{2}}\hbar
+\Lambda_{2}\kappa_{2_{-}}\sqrt{\frac{\hbar}{m\omega_{2}}}-i\Lambda_{1}%
\lambda_{2_{-}}\sqrt{\frac{\hbar}{m\omega_{1}}}\right)  }{2\sinh[\Gamma
](l_{-}\hbar\cos[\Upsilon]-F_{+}\sin[\Upsilon])+\cosh[\Gamma](2l_{+}\hbar
\sin[\Upsilon]-G_{-}\cos[\Upsilon])+G_{+}},\label{xi2}%
\end{equation}
where
\begin{align}
f_{i} &  =-\frac{m\hbar\omega_{i}}{4}\left(  \frac{1}{m}\mu^{2}+\frac
{m\omega_{j}^{2}\theta_{ji}^{2}}{4\nu^{2}\hbar^{2}}\right)  +\frac{\hbar
}{4m\omega_{i}}\left(  m\omega_{i}^{2}\nu^{2}+\frac{\varsigma_{ji}^{2}}%
{4m\mu^{2}\hbar^{2}}\right) \label{angulos} ,\\
g_{i} &  =\frac{m\hbar\omega_{i}}{2}\left(  \frac{1}{m}\mu^{2}+\frac
{m\omega_{j}^{2}\theta_{ji}^{2}}{4\nu^{2}\hbar^{2}}\right)  +\frac{\hbar
}{2m\omega_{i}}\left(  m\omega_{i}^{2}\nu^{2}+\frac{\varsigma_{ji}^{2}}%
{4m\mu^{2}\hbar^{2}}\right)  ,\\
l_{\pm} &  =\frac{1}{2\hbar}\left[  \sqrt{\frac{\omega_{1}}{\omega_{2}}%
}\left(  \frac{1}{2m}\varsigma_{12}+\frac{1}{2}m\omega_{2}^{2}\theta
_{12}\right)  \pm\sqrt{\frac{\omega_{2}}{\omega_{1}}}\left(  \frac{1}%
{2m}\varsigma_{12}+\frac{1}{2}m\omega_{1}^{2}\theta_{12}\right)  \right]  ,\\
\lambda_{2\pm} &  =(\cosh[r_{2}]\pm e^{i\phi_{2}}\sinh[r_{2}])\left(
\sin\left[  \frac{\Upsilon}{2}\right]  \cosh\left[  \frac{\Gamma}{2}\right]
\pm\cos\left[  \frac{\Upsilon}{2}\right]  \sinh\left[  \frac{\Gamma}%
{2}\right]  \right)  ,\\
\kappa_{2\pm} &  =(\cosh[r_{2}]\mp e^{i\phi_{2}}\sinh[r_{2}])\left(
\cos\left[  \frac{\Upsilon}{2}\right]  \cosh\left[  \frac{\Gamma}{2}\right]
\pm\sin\left[  \frac{\Upsilon}{2}\right]  \sinh\left[  \frac{\Gamma}%
{2}\right]  \right)  ,
\end{align}
and $F_{-}=f_{1}-f_{2}$, $F_{+}=f_{1}+f_{2}$, $G_{-}=g_{1}-g_{2}$, and
$G_{+}=g_{1}+g_{2}$. With the eigenvalues given by

%

\begin{equation}
\varepsilon_{n_{1},n_{2}} = \bar{C} \left(  n_{1}+ \frac{1}{2}\right)  +
\bar{D} \left(  n_{2} + \frac{1}{2} \right)  -\left(  \bar{C} | \chi_{1} |^{2}
+\bar{D} | \chi_{2} |^{2}-\bar{E}\right)  , \label{energies}%
\end{equation}
where
\begin{align}
\bar{C}  &  = \frac{1}{2} \operatorname{sech} [ 2 r_{1}] [ G_{+} \cosh[
\Gamma] - 2 \sin[\Upsilon] ( l_{+} \hbar+ F_{+} \sinh[ \Gamma]) + \cos[
\Upsilon] ( G_{-} + 2 l_{-} \hbar\sinh[ \Gamma])],\nonumber\\
\bar{D}  &  = \frac{1}{2} \operatorname{sech} [ 2 r_{2}] [ ( G_{+} + 2 \sin[
\Upsilon] (l_{+} \hbar\cosh[ \Gamma] - F_{+} \sinh[ \Gamma]) - \cos[ \Upsilon]
( G_{-} \cosh[ \Gamma] - 2 l_{-} \hbar\sinh[ \Gamma]) \nobracket],\nonumber\\
\bar{E}  &  =\Omega_{1}+\Omega_{2} - \frac{1}{2} \left[  g_{1} \cos^{2}
\left[  \frac{\Upsilon}{2} \right]  +g_{2} \sin^{2} \left[  \frac{\Upsilon}{2}
\right]  - \hbar l_{+} \sin[ \Upsilon] \right]  ( \cosh[\Gamma] - 1).
\label{last}%
\end{align}
The analytical solution allows us to analyze the effects of the
non-Hermiticity and noncommutativity in the Hamiltonian eigenstates and
eigenenergies. Through the equations (\ref{angulos}-\ref{last}) we can easily
see that the functions $\Upsilon$, $\Gamma$ and $r_{i}$ are dependent on the
noncommutative parameters and independent on the non-Hermitian parameters,
and this implies that the rotation and the compression in the states are
independent on the non-Hermiticity of the Hamiltonian. In fact we can analyze
only the effects of the noncommutativity, by doing $\{\omega_{i}, \gamma
_{i}\}\in\mathbb{R}$, $\delta=-i\tilde{\delta}$, this leads $V_{i}=\gamma_{i}$
and $T_{i}=\tilde{\delta}_{i}$, the functions $\Lambda_{i}$, $\Xi_{i}$ and
$\Omega_{i}$ become $\Lambda_{i}=\left(  \frac{\varsigma_{ji} \gamma_{j}}{2
\mu\hbar} + \nu\tilde{\delta}_{i} \right)  $, $\Xi_{i}=\left(  \mu\gamma_{i}
-\frac{\theta_{\operatorname{ji} } \tilde{\delta}_{j}}{2 \nu\hbar} \right)  $,
$\Omega_{i}=\frac{\tilde{\delta}_{i}^{2}}{m\omega_{i}^{2}}$. The only change
in relation to the eigenstates (\ref{eigenstate}) is a change in the
displacemnt operator $\chi_{i}$, eqs. (\ref{xi1}-\ref{xi2}). Clearly, the
deformation in the space given by noncommutativity, results in the rotation
of the state, in the compression of the modes and contributes to the
displacement. The only change in the eigenenergies (\ref{energies}) is in the
quadratic terms dependent on the displacement parameter. This shows that most
of the contribution in the evolution of this state and in the eigenenergies
comes from the noncommutativity. To see the contribution of non-Hermiticity in
this Hamiltonian we take $\{\omega_{i}, \gamma_{i}, \delta{i}\}\in\mathbb{R}$,
$\theta_{ij}=\zeta_{ij}\rightarrow0$ and $\mu=\nu\rightarrow1$, meaning that
the Hamiltonian (\ref{11}) goes to be ``commutative" (described by
conventional quantum mechanics) and non-Hermitian with the eigenvectors given
by
\begin{equation}
|\psi\rangle=D_{1}(-\chi_{1})D_{2}(-\chi_{2})|n_{1},n_{2}\rangle,
\end{equation}
where
\begin{equation}
\chi_{i} =\sqrt{\frac{1}{2m\hbar\omega^{3}_{i}}}\delta_{i}\tan\left[
\theta_{B_{i}}\right]  +i\sqrt{\frac{m}{2\omega_{i}\hbar}}\gamma_{i} ,
\label{NHD}%
\end{equation}
with the eigenvalues
\begin{equation}
\varepsilon_{n_{1},n_{2}}= \left(  n_{1}+\frac{1}{2}\right)  \omega_{1}%
\hbar+\left(  n_{2}+\frac{1}{2}\right)  \omega_{2}\hbar+\frac{1}{2m}\left(
\frac{\delta_{1}^{2}}{\omega_{1}^{2}}+\frac{\delta_{2}^{2}}{\omega_{2}^{2}%
}\right)  . \label{NHE}%
\end{equation}
Exactly as we had predicted, the contribution of non-Hermiticity occurs in the
displacement parameter proportional to $\gamma_{i}$, adding an energy to the
system proportional to the non-Hermitian parameter $\delta_{i}$. We can
compare the energies, (\ref{energies}) and (\ref{NHE}), with the energy of the
Hermitian linearly amplified harmonic oscillator. For that we just have to
take $\{\omega_{i}, \gamma_{i}\}\in\mathbb{R}$, $\delta=-i\tilde{\delta}$,
$\theta_{ij}=\zeta_{ij}\rightarrow0$ and $\mu=\nu\rightarrow1$, meaning that
the Hamiltonian (\ref{fullHHNC}) goes to be ``commutative" (described by
standard quantum mechanics) and Hermitian with the eigenvectors given by
\begin{equation}
|\psi\rangle=D_{1}(-\chi_{1})D_{2}(-\chi_{2})|n_{1},n_{2}\rangle, \label{SHH}%
\end{equation}
where
\begin{equation}
\chi_{i}=-\sqrt{\frac{1}{2m\omega_{i}^{3}\hbar}}\tilde{\delta}_{i}+i
\sqrt{\frac{m}{2\omega_{i}\hbar}}\gamma_{i}, \label{DHH}%
\end{equation}

and the eigenvalues
\begin{equation}
\varepsilon_{n_{1},n_{2}}=\left(  n_{1}+\frac{1}{2}\right)  \omega_{1}+\left(
n_{2}+\frac{1}{2}\right)  \omega_{2}-\frac{1}{2m}\left(  \frac{\tilde{\delta
}_{1}^{2}}{\omega_{1}^{2}}+\frac{\tilde{\delta}_{2}^{2}}{\omega_{2}^{2}%
}\right)  . \label{HE}%
\end{equation}
As can be seen from the equation (\ref{HE}), the linear amplification in the
Hermitian Hamiltonian draws energy from the system in proportion to the square
of the amplification parameters, $\gamma_{i}$ and $\delta_{i}$, and the state
of the system is a displacement of the number states, with that displacement
proportional to $\gamma_{i}$ and to $-\tilde{\delta_{i}}$. We can choose the
free parameter on the Dyson map, $\theta_{B_{i}}$, such that the displacement
parameter in the state whose evolution is given by a non-Hermitian Hamiltonian
(\ref{NHD}) stays equal to the displacement parameter of the state of their
Hermitian Hamiltonian (\ref{DHH}). Thus, both states are the same. Although we
choose $\theta_{B_{i}}$, so that both have the same states, the Hamiltonians
remain different and this results in different eigenenergies. Moreover, even
with the isospectral partner of the non-Hermitian Hamiltonian obtaining the
same states of the Hermitian Hamiltonian, the latter has a smaller energy,
eq.(\ref{HE}), compared to the isospectral partner of the non-Hermitian
Hamiltonian, eq.(\ref{NHE}).

As mentioned in section \ref{sec3} (see equations (\ref{eq10}) and (\ref{eq11})), our method allows to obtain the Wigner function associated to the Hamiltonian of the system, eq.(\ref{11}) or for the Hermitian commutative Hamiltoninan resulting from the application of the Dayson and Seiberg-Witten maps. However, we do not write the expression for the Wigner function here because it is mathematically cumbersome and no extra physical information is obtained from it. Regardless this, we reinforce that our method has been applied and physical information acquired when applying the Dyson and Seiberg-Witten maps on the Hamiltonian (\ref{11}). 

\section{Conclusions}

\label{con}

In this work we have formulated a unified formalism to treat non-Hermitian and
noncommutative Hamiltonians. By considering the Dyson and Seiberg-Witten maps
for the non-Hermitian and noncommutative operators respectively, it was
possible in the phase-space formalism of the quantum mechanics to develop a
formal and correct way to employ both generalizations in the canonical quantum mechanics.

Since the non-Hermitian aspects of quantum mechanics had been elucidated in
many quantum systems as mentioned in the introduction, we ask how these
features will appear in a scenario where the noncommutative of the
phase-space becomes relevant. For this, the formalism developed in this work
presents the correct way to deal simultaneously with both aspects of the
quantum generalizations involved. Our example, even in the particular case,
allows to show the applicability of the method and its practical consequences.
Furthermore, in a future work, we intend to analyze the complete Hamiltonian in
appendix in details, on the aspects of the Wigner function and the
eigenenergies and its full properties.

We would like to compare our method to that presented in \cite{Roy}. In that
work, the authors address the problem from a different point of view. The main
point of analysis is to impose $\mathcal{PT}$-symmetry on the general
commutation relations of the noncommutative quantum mechanics, eq.(\ref{2})
in our text. The authors then verify that this realization does not preserve
the commutation relations $\mathcal{PT}$-symmetric. To solve the problem, they
propose another form for the relations, for instance, they write $\lbrack
q_{k},q_{\ell}]=\theta_{k\ell}$, in our notation. Although the authors argue
that this procedure solve the problem, it is direct to note that in doing so
the relations are no longer invariant under self-adjoint transformation. In
order to avoid this apparent problem we follow another direction in our work,
explicitly showing that a robustness way is, given a non-Hermitan and
noncommutative operator, first obtain the Hermitian counterpart of the
operator and then perform the Siberg-Witten map, not being necessary change
the commutation relations. Another different approach has been taken in
\cite{Andreas01}, where is argued about the possibility of the $\mathcal{PT}%
$-symmetry change the sign of the new noncommutative constants in the theory, and other references that discuss $\mathcal{PT}$-symmetry and deformation in the Heisenberg algebra are \cite{new01, new02}. Again, our method does not assume this option, mainly based on the fact that
these constants are invariant such that the Planck constant. 
\begin{flushleft}
{\Large \textbf{Acknowledgements}}
\end{flushleft}

JFGS would like to thank CAPES (Brazil) and Federal University of ABC for
support, FSL would like to thank CNPq (Brazil) for support Grant No.
150879/2017-2, OSD would like to thank CNPq (Brazil) for financial support
through the Grant No. 153119/2018-7, and MHYM would like to thank CAPES
(Brazil) for support and the City University London for kind hospitality. The
authors are in debt to Prof. A. Fring of the City University London, for many
fruitful discussions.

\section*{Appendix}

\label{app}

Here it will be presented the general case of the Hamiltonian in example
above. After we use the Dyson map in the Hamiltonian $\mathcal{H}_{NH,
NC}^{\mathcal{PT}}(\mathbf{q}, \mathbf{p})$ we get, {\small
\begin{align}
H_{H, NC}(\mathbf{q}, \mathbf{p})  &  = \sum_{i = 1}^{\mbox{mod2}} [\alpha
_{i}\, p_{i}^{2} + \beta_{i}\, q_{i}^{2} + p_{i} \,(2i\alpha_{i}\, (\hbar
A_{i} + \zeta_{i+1}B_{i+1, i}) + \gamma_{i}) + iq_{i}(2\beta_{i}\,
(\theta_{i+1, i} A_{i+1} - \hbar B_{i}) + \delta_{i})\nonumber\\
&  - \alpha_{i}\, (\hbar A_{i} + \zeta_{i+1, i}B_{i+1})^{2} - \beta
_{i}\,(\theta_{i+1,i}A_{i+1} - \hbar B_{i})^{2}\nonumber\\
&  + i\gamma_{i}(\hbar A_{i} + \zeta_{i+1, i} B_{i+1}) + \delta_{i}(\hbar
B_{i} - \theta_{i+1, i}A_{i+1})],
\end{align}
where mod2 implies that $2+1 = 1$. }

{\small After the transformation, the Hamiltonian $H_{H, NC}(\mathbf{q},
\mathbf{p})$ must be Hermitian, i. e., $H_{H, NC}(\mathbf{q}, \mathbf{p}) =
H_{H, NC}(\mathbf{q}, \mathbf{p})^{\dagger}$. Using this condition, the polar
notation for complex quantities and after some mathematical manipulations, one
has, }{\footnotesize
\begin{align}
H_{H, NC} ( q, p)  &  = \sum_{i = 1}^{2} \left[  \alpha_{i} p_{i}^{2} +
\beta_{i} q_{i}^{2} + p_{i} \left(  - \frac{\alpha_{i} \varsigma_{ji}
\hbar\delta_{j}( \tan[ \theta_{B_{j}}] - \tan[\theta_{A_{i}}])}{\beta_{j} (
\varsigma_{ji} \theta_{ij} +\hbar^{2})} + \gamma_{i} \right)  + q_{i} \left(
\frac{\delta_{i}( \tan[\theta_{A_{j}}] \theta_{ji} \varsigma_{ij} +
\tan[\theta_{B_{i}}] \hbar^{2})}{( \theta_{ji} \varsigma_{ij} +\hbar^{2})}
\right)  \right. \nonumber\\
&  \left.  - \alpha_{i} \left(  \frac{i \delta_{j} ( \tan[ \theta_{B_{j}}] -
\tan[ \theta_{A_{i}}]) \varsigma_{ji} \hbar}{2 \beta_{j} (\varsigma_{ji}
\theta_{ij} + \hbar^{2})} \right)  ^{2} - \frac{\delta_{i}^{2}}{4 \beta_{i}}
\left(  - 1 - \frac{i ( \tan[ \theta_{A_{j}}]\theta_{ji} \varsigma_{ij} +
\tan[ \theta_{B_{i}}] \hbar^{2})}{(\theta_{ji} \varsigma_{ij} + \hbar^{2})}
\right)  ^{2} \right. \nonumber\\
&  \left.  - \gamma_{i}\delta_{j} \left(  \frac{( \tan[ \theta_{B_{j}}] -
\tan[ \theta_{A_{i}}]) \varsigma_{ji} \hbar}{2 \beta_{j} ( \varsigma_{ji}
\theta_{ij} + \hbar^{2})} \right)  + \frac{\delta_{i}^{2}}{2 \beta_{i}}
\left(  1 + i \frac{( \tan[ \theta_{B_{i}}] \hbar^{2} + \tan[ \theta_{A_{j}}]
\theta_{ji} \varsigma_{ij})}{( \theta_{ji} \varsigma_{ij} + \hbar^{2})}
\right)  \right]  , \label{0001}%
\end{align}
where $j = 1+i\mbox{mod}2$. If we consider the parameters of the Dyson map
real we have $\theta_{B_{i}} = \theta_{A_{i}} = 0$, and this leads the
equation above for
\begin{equation}
\mathcal{H}_{H, \operatorname{NC} }^{} ( q, p) = \sum_{i = 1}^{2} \left[
\alpha_{i} p_{i}^{2} + \beta_{i} q_{i}^{2} + p_{i} \gamma_{i} + \frac
{\delta_{i}^{2}}{4 \beta_{i}} \right]  . \label{eq141}%
\end{equation}
}

{\footnotesize Now, using the Seiberg-Witten map in the Hamiltonian (\ref{0001}) one
has, after mathematical manipulations,
\begin{equation}
\mathrm{H}_{H, C}(\mathbf{Q}, \mathbf{P}) = \sum_{i = 2}^{2} \, F_{i}%
\,P_{i}^{2} + G_{i}\, Q_{i}^{2} + H_{i}\{P_{i}, Q_{i}\} + I_{i}\, P_{i} +
K_{i}\,Q_{i} + L_{i},
\end{equation}
where the constants $F_{i}$, $G_{i}$, $H_{i}$, $I_{i}$, $K_{i}$ and $L_{i}$
are given by
\begin{align}
F_{i}  &  = \alpha_{i} \mu^{2} + \frac{\beta_{j} \theta_{ji}^{2}}{4 \nu^{2}
\hbar^{2}},\nonumber\\
G_{i}  &  = \beta_{i} \nu^{2} + \frac{\alpha_{j} \varsigma_{ji}^{2}}{4 \mu
^{2}\hbar^{2}},\nonumber\\
H_{i}  &  = \frac{\alpha_{i} \varsigma_{\operatorname{ij} }}{2 \hbar} -
\frac{\beta_{j}\theta_{ji}}{2 \hbar},\nonumber\\
I_{i}  &  = \gamma_{i} - \frac{\mu\delta_{j}}{( \varsigma_{ji}\theta
_{\operatorname{ij} } + \hbar^{2})} \left(  \frac{\alpha_{i}\varsigma_{ji}
\hbar}{\beta_{j}} ( \tan[ \theta_{B_{j}}] - \tan[ \theta_{A_{i}}])
+\frac{\theta_{ji} }{2 \nu\hbar} ( \tan[ \theta_{A_{i}}]\varsigma_{ji}
\theta_{\operatorname{ij} } + \tan[ \theta_{B_{j}}] \hbar^{2})\right)
,\nonumber\\
K_{i}  &  = \frac{\varsigma_{ji}}{2 \mu\hbar} \gamma_{j} + \frac{\delta_{i}%
}{(\theta_{ji} \varsigma_{\operatorname{ij} } + \hbar^{2})} \left(  -
\frac{\alpha_{j}\varsigma_{ji} \varsigma_{\operatorname{ij} } \hbar( \tan[
\theta_{B_{i}}] - \tan[ \theta_{A_{j}}])}{2 \mu\hbar\beta_{i}} + \nu( \tan[
\theta_{A_{j}}]\theta_{ji} \varsigma_{\operatorname{ij} } + \tan[
\theta_{B_{i}}] \hbar^{2})\right)  ,\nonumber\\
L_{i}  &  = - \alpha_{i} \left(  \frac{i \delta_{j} ( \tan[ \theta_{B_{j}}] -
\tan[\theta_{A_{i}}]) \varsigma_{ji} \hbar}{2 \beta_{j} (\varsigma_{ji}
\theta_{\operatorname{ij} } + \hbar^{2})} \right)  ^{2} - \beta_{i} \left(  -
\frac{\delta_{i}}{2 \beta_{i}} - \frac{i ( \tan[ \theta_{A_{j}}]\theta_{ji}
\varsigma_{\operatorname{ij} } + \tan[ \theta_{B_{i}}] \hbar^{2}) \delta_{i}%
}{2 \beta_{i} ( \theta_{ji} \varsigma_{\operatorname{ij} } + \hbar^{2})}
\right)  ^{2}\nonumber\\
&  + i \gamma_{i} \left(  + \frac{i ( \tan[ \theta_{B_{j}}] - \tan
[\theta_{A_{i}}]) \delta_{j} \varsigma_{ji} \hbar}{2 \beta_{j} (\varsigma_{ji}
\theta_{\operatorname{ij} } + \hbar^{2})} \right)  + \delta_{i} \left(
\frac{\delta_{i}}{2 \beta_{i}} + i \frac{\delta_{i} ( \tan[ \theta_{B_{i}%
}]\hbar^{2} + \tan[ \theta_{A_{j}}] \theta_{ji} \varsigma_{\operatorname{ij}
})}{2\beta_{i} ( \theta_{ji} \varsigma_{\operatorname{ij} } + \hbar^{2})}
\right)  ,\nonumber
\end{align}
}

{\footnotesize It is direct to see that this Hamiltonian naturally differs
from the one just with noncommutative effects. In a next work we intend to
treat mathematically and physically this complex Hamiltonian in details,
analyzing its eigenstates and eigenenergies. }

\end{document}